\documentclass[11pt,twoside]{article}

\usepackage{asp2006}
\usepackage{epsf}
\usepackage{psfig}

\begin{document}
\title{Models of the Intergalactic Gas in Stephan's Quintet}   
\author{Jeong-Sun Hwang$^1$, Curtis Struck$^1$, Florent Renaud$^{2,3}$, and Philip Appleton$^4$}   
\affil{$^1$Department of Physics and Astronomy, Iowa State University, Ames, IA 50011, USA}    
\affil{$^2$Observatoire Astronomique and CNRS UMR 7550, Universit\'e de Strasbourg, 11 rue de l'Universit\'e, F-67000 Strasbourg, France} 
\affil{$^3$Institut f\"ur Astronomie der Univ. Wien, T\"urkenschanzstr. 17, A-1180 Vienna, Austria} 
\affil{$^4$NASA Herschel Science Center (NHSC), California Institute of Technology, Mail code 100-22, Pasadena, CA 91125, USA} 

\begin{abstract} 
We use smoothed particle hydrodynamics (SPH) models to study the large-scale morphology 
and dynamical evolution of the intergalactic gas in Stephan's Quintet, 
and compare to multiwavelength observations. 
Specifically, we model the formation of the hot X-ray gas, the large-scale shock, 
and emission line gas as the result of NGC 7318b colliding with the group. 
We also reproduce the N-body model of Renaud and Appleton 
for the tidal structures in the group.
\end{abstract}



\section{Introduction}
Stephan's Quintet (hereafter SQ) is a compact group of galaxies discovered 
by \'Edouard Stephan in 1877. 
Since it is relatively close and shows many interesting features of galaxy-galaxy and galaxy-intergalactic medium (IGM) interactions SQ has been a popular object of study. 
Its multiple interactions, however, make difficult to know its dynamical history 
and interpret some observed aspects. 
In order to study SQ's dynamical evolution and the cause of the large-scale shock structure in its IGM seen in multiwavelength observations we have performed numerical simulations of SQ.
In this proceedings we first review SQ's morphological and kinematical structures that we tried to simulate and then present our preliminary model results. Finally we discuss some adjustment of our current model to obtain better results.

\section{Past Study of SQ}
The optical morphology of SQ is shown in Figure 1. NGC 7320 is a foreground galaxy which has a considerably smaller redshift than others \citep{BB1961}. Assuming a group distance of 94 Mpc and a Hubble constant of 70 km s$^{-1}$ Mpc$^{-2}$, the group except NGC 7318b has a redshift about 6600 km s$^{-1}$ and NGC 7318b is coming toward us with the relative velocity of about 900 km s$^{-1}$. Two long tidal tails extending from NGC 7319 toward NGC 7320c are visible. The southern tail, which is fainter, passes behind the foreground galaxy NGC 7320 and runs almost parallel with the northern tail. 
NGC 7319 is known to be a Seyfert 2 galaxy.

The VLA HI observations reveal that most of the HI gas resides outside of the galaxies \citep{Williams2002}. The observed HI clouds along the two optical parallel tails are thought to be the result of one or more past encounters of NGC 7320c with the group \citep{MSM1997, Sulentic2001}. The past encounters might have stripped and removed most of the gas from NGC 7319 to the intergalactic space. 
NGC 7318b is colliding with the IGM which has formed during the past encounters \citep{Sulentic2001, Xu2003}. This ongoing collision may produce the large-scale shock in the IGM observed in radio continuum \citep{AH1972, vdHR1981}, X-rays \citep{Pietsch1997, Trinchieri2003, O'Sullivan2009} and mid-IR H$_{2}$ emission \citep{Appleton2006}.

\begin{figure}[!ht]
\centering%
\epsscale{0.55}\plotone{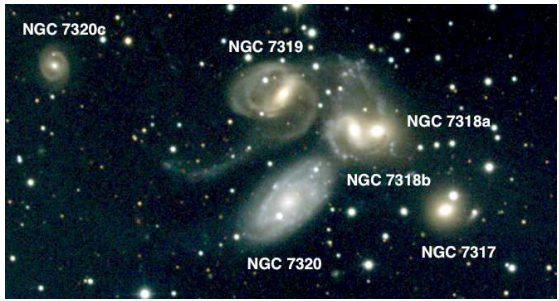}
\caption{The optical morphology of SQ. North is up and east to the left. Credit: NOAO/AURA/NSF}
\end{figure}

\section{Numerical Simulations of SQ}
Renaud and Appleton had performed N-body simulations of SQ using mostly NEMO, a stellar dynamics toolbox \citep{Teuben1995}. Considering observational constraints they deduced probable ranges of initial conditions for their model galaxies and reproduced SQ's general morphology \citep{RA2009}. 
In order to add thermohydrodynamical effects to the N-body model and study dynamical evolution of SQ we have produced an SPH code by modifying the SPH code of \citet{Struck1997} which was designed for studying collisions between galaxy pairs (see \citealp{Struck1997} for details). We performed the simulation with four strongly interacting members. Each of our model galaxies has disks with gas and collisionless star particles and a rigid dark matter halo. We took the initial parameters from the N-body model and adjusted them (mainly initial velocities of model galaxies) to use in the SPH model which has more extended dark matter halo potentials. We used total 81120 (40320 gas + 40800 star) particles and the mass ratio of 1(NGC 7319) : 0.7(NGC 7318a) : 0.6(NGC 7318b) : 0.2(NGC 7320c).

Our preliminary results of the SPH simulation are as follows. Figure 2.a and b show the production of two long tidal tails at an earlier time step. Since the effect of NGC 7318b to the group is small at early times we have put NGC 7318b into the simulation at this time step. Figure 2.c and d represent current morphology of SQ and the collision between NGC 7318b and the IGM from the model. 

\begin{figure}[!ht]
\centering%
\epsscale{1.0}\plotone{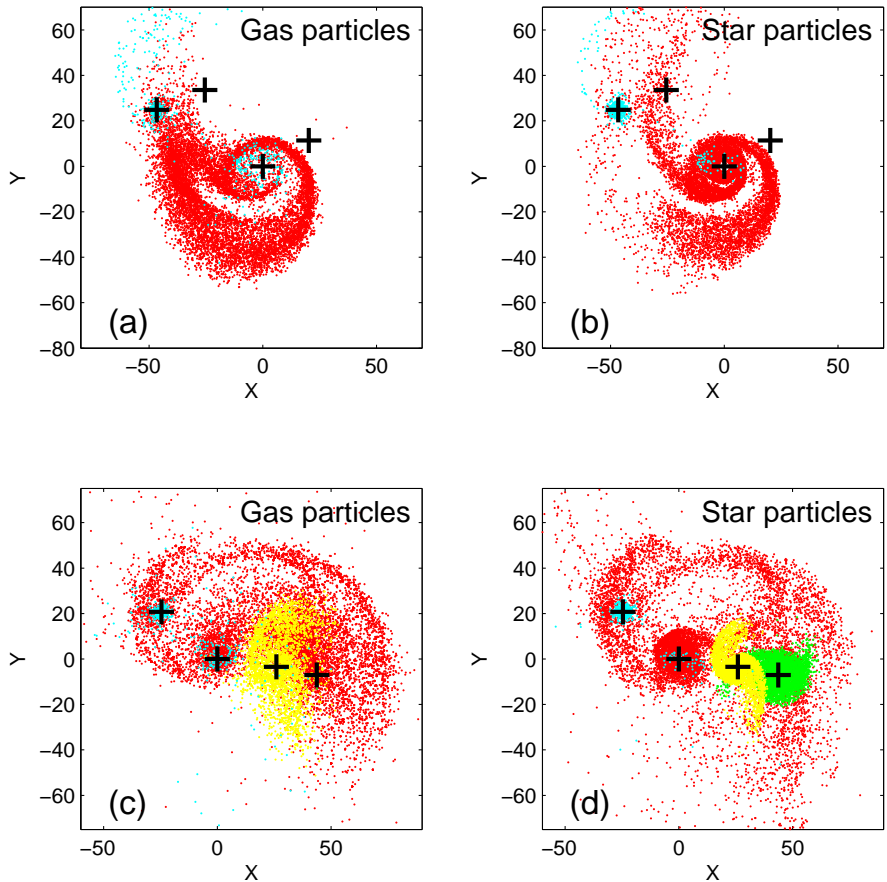}
\caption{Snapshots of the model galaxies at an early time step (top panels) 
and time near the present (bottom panels). 
The left and right panels show the distributions of gas and star particles respectively. 
Particles originating from NGC 7319, NGC 7318a, NGC 7318b, and NGC 7320c are shown with red, green, yellow, and cyan dots respectively.
The center positions of four galaxies are marked with crosses. 
{\it (a) and (b)}: Particles originating from NGC 7319 and NGC 7320c are shown for simplicity. 
The two long parallel tidal tails have developed. 
The (x, y, z) values of the centers of NGC 7319, NGC 7318a, NGC 7318b and NGC 7320c 
are (0, 0, 0), (-26, 34, 24), (20, 11, -98), and (-47, 25, -24) respectively.
{\it (c) and (d)}: Particles originating from all four galaxies are shown. NGC 7318b is colliding with the group. 
The (x, y, z) values of the centers of NGC 7319, NGC 7318a, NGC 7318b and NGC 7320c 
are (0, 0, 0), (44, -7, -22), (26, -3, 12), and (-25, 21, 4) respectively.}
\end{figure}
 
\section{Discussion and Future Work}
The general morphology of SQ was reproduced except the position of the southern tail and inner structures of spiral members. With our extended halo potentials the southern tail was placed too far out, so NGC 7318b did not hit many particles in the tidal tails. 
We will adjust it to make a more direct collision between the galaxy and the IGM. 
Also, to obtain well-matched results in radial velocities and temperatures of gas particles (not shown in this article) with those of HI, radio continuum, and X-ray observations, we will be working on Òfine tuningÓ the initial parameters and cooling in the models.  

\acknowledgements 
We thank the conference organizers for providing an opportunity to discuss this topic and financial support.
This work has been supported by NASA Chandra grant AR90010B.


%
\end{document}